\documentclass[preprint,12pt]{elsarticle}

\usepackage{amssymb}
\usepackage{color}

\journal{}

\begin{document}

\begin{frontmatter}



\title{Coupled effects of epidemic information and risk awareness on contagion}


\author {Wen-Juan Xu$^a$}
\author{Chen-Yang Zhong$^b$}
\author {Hui-Fen Ye$^c$}
\author {Rong-Da Chen$^d$}
\author {Tian Qiu$^e$}
\author {Fei Ren$^f$}
\author{Li-Xin Zhong$^d$}\ead{zlxxwj@163.com}

\address[label1]{School of Law, Zhejiang University of Finance and Economics, Hangzhou, 310018, China}
\address[label2]{Department of Statistics, Stanford University, Stanford, CA 94305-4065, USA}
\address[label3]{School of Accounting, Zhejiang University of Finance and Economics, Hangzhou, 310018, China}
\address[label4]{School of Finance and Coordinated Innovation Center of Wealth Management and Quantitative Investment, Zhejiang University of Finance and Economics, Hangzhou, 310018, China}
\address[label5]{School of Information Engineering, Nanchang Hangkong University, Nanchang, 330063, China}
\address[label6]{School of Business and Research Center for Econophysics, East China University of Science and Technology, Shanghai, 200237, China}

\begin{abstract}
In modern society, on the one hand, a highly developed transportation system has greatly promoted population mobility, which makes the prevention and control of an epidemic difficult. On the other hand, a highly developed information system has promoted real-time remote communication, which helps people obtain timely and accurate epidemic information and protect themselves from being infected. In order to make best use of the advantages and bypass the disadvantages of modern technologies in the prevention and control of an infectious disease, there is a need to give an insight into the relationship between the spread of an epidemic and people's risk aversion behaviors. By incorporating delayed epidemic information and self-restricted travel behavior into the SIS model, we have investigated the coupled effects of timely and accurate epidemic information and people's sensitivity to the epidemic information on contagion. In the population with only local random movement, whether the epidemic information is delayed or not has no effect on the spread of the epidemic. People's high sensitivity to the epidemic information leads to their risk aversion behavior and the spread of the epidemic is suppressed. In the population with only global person-to-person movement, timely and accurate epidemic information helps an individual cut off the connections with the infected in time and the epidemic is brought under control in no time. A delay in the epidemic information leads to an individual's misjudgment of who has been infected and who has not, which in turn leads to rapid progress and a higher peak of the epidemic. In the population with coexistence of local and global movement, timely and accurate epidemic information and people's high sensitivity to the epidemic information play an important role in curbing the epidemic. A theoretical analysis indicates that people's misjudgment caused by the delayed epidemic information leads to a higher encounter probability between the susceptible and the infected and people's self-restricted travel behavior helps reduce such an encounter probability. A functional relation between the ratio of infected individuals and the susceptible-infected encounter probability has been found.
\end{abstract}


\end{frontmatter}


\section{Introduction}
\label{sec:introduction}
More recently, the world has witnessed a devastating outbreak of COVID-19\cite{ventura1,johnson1,velasquez1,ribeiro,james1}. Compared with the spread of the plagues in early human history\cite{bootsma1,hatchett1}, the spread of COVID-19 has a greater average daily dispersion. Such a difference results from the development of modern transit system, which makes the people around the world highly connected\cite{zhong1,manrique1,zheng2,ruan1}. Therefore, from the view point of the prevention and control of an infectious disease, today's mobile society has brought us more challenges than ever before.

Traditionally, spatial isolation is an effective method for the prevention and control of an infectious disease. However, in modern society, long-term isolation means that people may lose their jobs on which they rely for survival. In addition to that, long-term isolation may also cause people's psychological problems\cite{ferguson1,brauer1}. Under such circumstances, how to keep people from being infected and maintain appropriate outdoor activities at the same time has become one of the major problems which need to be investigated in depth.

With the help of highly developed communication technologies, such as the Internet and the mobile phone, people are easy to get timely and accurate epidemic information. Such an advantage provides people with more opportunities to find effective measures for epidemic suppression in addition to spatial isolation\cite{stauffer1,schweitzer1,lorenz1,pfitzner1,scholtes1}. Therefore, from the view point of the prevention and control of an infectious disease, today's information society has brought us  more opportunities than ever before\cite{perc1,kumar1,guo1,ribeiro1,khajanchi1}.

A variety of network models have been borrowed to analyze the coupling of information diffusion and epidemic spreading in the last two decades\cite{zheng1,xu1,wang1,roni1,pastor1,hu1,wu1}. In a dynamical network, people can effectively avoid being infected by cutting off the connections with the infected and relinking to the uninfected. In a multilayer network, the diffusion of epidemic information and the spread of an infectious disease are coupled together with common nodes. The topological properties of the information network and the epidemic transmission network have a great impact on the spread of the epidemic. In a metapopulation network, people's moving patterns affect their encounter probability, which finally determines the speed and scope of the spread of an epidemic\cite{min1,zhong2,moorel,granell1,funk1}.

Although the studies related to the coupling of information diffusion and epidemic spreading have investigated the effects of communication structures and interaction structures on the spread of an infectious disease, in order to give an exact answer to the question about how to keep people from being infected and maintain appropriate outdoor activities at the same time, we need to have a clear view of the time-delayed effect and the cognitive effect on the spread of the epidemic, which are still lack of in-depth discussions.

In the present work, by incorporating delayed epidemic information and self-restricted travel behavior into the SIS model, we have investigated the coupled effects of the timeliness and accuracy of epidemic information and people's sensitivity to the epidemic information on the spread of the epidemic. The following are our main findings.

(1) In the population with only local random movement, whether the epidemic information is delayed or not has little effect on the spread of an infectious disease. People's self-restricted travel behavior has a great impact on the spread of the epidemic. A higher level of sensitivity to the epidemic information leads to a lower level of going-out frequency into the mostly infected crowd and the spread of the epidemic is suppressed effectively.

(2) In the population with only global person-to-person movement, whether the epidemic information is delayed or not has a great impact on the spread of an infectious disease. The timely and accurate epidemic information helps an individual cut off the immediate connections with the infected in time and the epidemic dies out in a short time. The delayed epidemic information leads to an individual's misjudgement of who has been infected and who has not. The delay in cutting off the suspected-infected connections leads to a fast diffusion and a high peak of the epidemic. People's self-restricted travel behavior helps reduce the disadvantages resulting from the delayed epidemic information.

(3) A theoretical analysis indicates that the spread of an infectious disease is determined by the encounter probability between the suspected and the infected. The delayed epidemic information causes an increase in the suspected-infected encounter probability in the global person-to-person movement, which leads to a fast diffusion and a high peak of the epidemic. The information-dependent risk-aversion behavior causes a decrease in the suspected-infected encounter probability, which leads to effective suppression of the epidemic in both the local and global movement.

The rest of the paper is organized as follows. The SIS model with delayed epidemic information and local-global movement is introduced in section two. Section three presents simulation results and discussions. Section four gives a theoretical analysis. A conclusion is drawn in section five.

\section{The model}
\label{sec:model}
In the present model, there are three coupled evolutionary processes: people's self-restricted travel behavior, information diffusion and epidemic transmission. The mechanisms mainly concerned in the present work are the delayed epidemic information and the sensitivity of people's moving probability to local and global epidemic information. In the following, we introduce the above three evolutionary processes respectively.

\subsection{\label{subsec:levelA}People's self-restricted travel behavior}

People's travel behavior consists of two coupled moving patterns: local random movement and global person-to-person movement. In the local random movement, an individual $i$ moves randomly within a confined spatial area, called local area in the following, with probability $v^{ram}_i$. Each local area consists of a $l\times l$ area with an overlapping $l\times \Delta l$ area between two adjacent local areas. The moving probability $v^{ram}_i$ of individual $i$ in the local area is

\begin{equation}
v^{ram}_i=v_0(1-\frac{n_I}{n})^\alpha,
\end{equation}
in which $v_0$ is the initial moving probability, $n_I$ and $n$ are the number of infected individuals and the total number of population in the local area, $\alpha$ represents an individual's sensitivity to the local epidemic information. The total area of all the local areas combined is a $L\times L$ regular network with eight paths for each node, which is satisfied with the equation

\begin{equation}
L\times L=[m\times(l-\Delta l)]\times [m\times(l-\Delta l)].
\end{equation}
Therefore, there are total $m\times m$ number of local areas in the present model.

In the global person-to-person movement, an individual $i$ moves purposefully to an individual $j$ who has an immediate connection with him with probability $v^{pur}_i$. The globally connected network is a random network with average degree $\bar k$ for each node and a Poisson degree distribution $P(k)$. The moving probability $v^{pur}_i$ is satisfied with the equation

\begin{equation}
v^{pur}_i=v'_0(1-\frac{N_I}{N})^\beta,
\end{equation}
in which $v'_0$ is the initial moving probability, $N_I$ and $N$ are the number of infected individuals and the total number of population in the global area, $\beta$ represents an individual's sensitivity to the global epidemic information. In addition to the global information, an individual $i$ has also personal information about who has been infected and who has not among his immediate neighbors. If he obtains the information that an immediate neighbor has been infected, he will cut off the connection with him.

The local movement and the global movement are coupled together with timescale $\tau$. For $\tau=0$, there is only local movement. For $\tau=1$, there is only global movement. For $0<\tau<1$, local movement and global movement coexist.

\subsection{\label{subsec:levelB}Information diffusion}

There are three kinds of epidemic information in the present model: the local epidemic information $\rho^{loc}_I=\frac{n_I}{n}$, the global epidemic information $\rho^{glo}_I=\frac{N_I}{N}$ and the person-to-person epidemic information $S-I$. If the epidemic information can be obtained in time, each individual determines his travel behavior according to the latest epidemic information, i.e. $\rho^{loc}_I(t)=\frac{n_I(t)}{n}$, $\rho^{glo}_I(t)=\frac{N_I(t)}{N}$ and $S-I(t)$. If the epidemic information is $\Delta$ time steps delay, each individual determines his travel behavior according to the delayed epidemic information, i.e. $\rho^{loc}_I(t)=\frac{n_I(t-\Delta)}{n}$, $\rho^{glo}_I(t)=\frac{N_I(t-\Delta)}{N}$ and $S-I(t-\Delta)$. Therefore, if the ratio of infected individuals in the population does not changes with time, that is $\frac{n_I(t)}{n}=\frac{n_I(t-\Delta)}{n}$ and $\frac{N_I(t)}{n}=\frac{N_I(t-\Delta)}{n}$, whether the local and global information are timely or not does not affect the local and global moving probability. But for person-to-person movement, if an individual was infected within the latest $\Delta-1$ steps, the information of $S-I(t)$ should be different from the information of $S-I(t-\Delta)$, which leads to an individual's misjudgement of who has been infected and who has not among his immediate neighbors. In such cases, the global person-to-person movement is quite possible to promote the spread of an infectious disease.

\subsection{\label{subsec:levelC}Epidemic transmission}

There are two routes for the spread of an infectious disease: group-to-person transmission and person-to-person transmission. In the local random movement, if a typical spot $(x, y)$ has been occupied by $n'_I$ infected individuals and $n'_S$ susceptible individuals, a susceptible individual $i$ in this spot will be infected with probability

\begin{equation}
P_i=\frac{n'_I}{n'_I+n'_S}P_I,
\end{equation}
in which $P_I$ is the infection probability of the infectious disease. An infected individual $j$ in this spot will become susceptible with probability

\begin{equation}
P_j=P_S,
\end{equation}
in which $P_S$ is the recovery probability of the infectious disease.

In the global person-to-person epidemic transmission, an individual $i'$ firstly randomly chooses an individual $j'$ from his immediate neighbors. If individual $i'$ is an infected individual, he becomes susceptible with probability $P_S$ and keeps infected with probability $1-P_S$. If individual $i'$ is a susceptible individual and individual $j'$ is an infected individual, individual $i'$ will be infected with probability $P_I$ on condition that individual $i'$ does not know that individual $j'$ has been infected. If  individual $i'$ knows that individual $j'$ has been infected, he will cut off the link and keeps susceptible.

\section{Simulation results and discussions}
\label{sec:results}
The epidemic information alerts people to the potential danger, which is possible to affect people's local and global travel behavior. The change of people's travel behavior would in turn lead to a change in the spread of the epidemic. In the coupling of people's self-restricted travel behavior and the spread of the epidemic, the timeliness and accuracy of epidemic information and people's attitude to such information are two key factors. In the following, we examine simulatively whether people's sensitivity to the risk information and the delay of the epidemic information affect the spread of the epidemic or not.

\begin{figure}
\includegraphics[width=12cm]{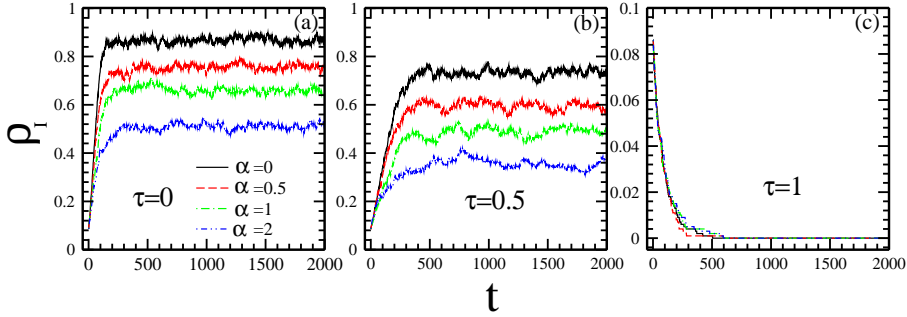}
\caption{\label{fig:epsart}Simulation of dynamic ratio of infected individuals $\rho_I$ for time t with timescale between global and local movement (a) $\tau=0$;  (b)$\tau=0.5$; (c)$\tau=1$. People's sensitivity to the epidemic information $\alpha$=0 (lines), 0.5 (slashes), 1 (slash dotted lines), 2 (slash dotted dotted lines). Other parameters are: $\alpha$=$\beta$, total population $N=1000$, average degree of each node in the random network $\bar k$=10, initial moving probability $v_0$=1, delayed time $\Delta$=0, infection rate $P_I=1$, recovery rate $P_S=0.01$, global area $L\times L=100\times 100$, local area $l\times l=11\times 11$, overlapping area $l\times\Delta l=11\times1$.}
\end{figure}

Firstly, suppose that people could get the epidemic information timely and accurately, we examine whether people's self-restricted travel behavior could suppress the spread of an infectious disease or not.

Figure 1 (a) shows that, for $\tau=0$, which corresponds to the situation where there is only local random movement, the time-dependent ratio of infected individuals $\rho_I$ is closely related to people's sensitivity to the risk information. For $\alpha$=$\beta$=0, which corresponds to the situation where people are insensitive to the epidemic information, within the range of $0<t<150$, $\rho_I$ increases quickly from $\rho_I\sim 0.1$ to $\rho_I\sim 0.86$. Within the range of $t>150$, $\rho_I$ fluctuates around $\rho_I\sim 0.86$ and the average value of $\rho_I$ changes little with the rise of $t$. For $\alpha$=$\beta$=0.5, which corresponds to the situation where people show low sensitivity to the epidemic information,
within the range of $0<t<300$, $\rho_I$ increases from $\rho_I\sim 0.1$ to $\rho_I\sim 0.76$. Within the range of $t>300$, $\rho_I$ fluctuates around $\rho_I\sim 0.76$ and the average value of $\rho_I$ changes little with the rise of $t$. A further increase in people's sensitivity to the epidemic information leads to a lower level of $\rho_I$ in the final steady state and the prolonged time to the final steady time. Such results indicate that, in the local random movement, an individual's self-restricted travel behavior can suppress the spread of an infectious disease effectively.

Figure 1 (b) shows that, for $\tau=0.5$, which corresponds to the situation where local movement and global movement coexist, the time-dependent ratio of infected individuals $\rho_I$ is also related to people's sensitivity to the risk information. Comparing the results in Figure 1 (a) with the results in Figure 1 (b), we find that, given the same value of people's sensitivity to the risk information, the coexistence of local movement and global movement leads to a lower level of $\rho_I$ in the final steady state and the prolonged time to the final steady time. Such results indicate that, in the coexistence of local movement and global movement, an individual's self-restricted travel behavior is helpful for inhibiting the spread of an infectious disease.

Figure 1 (c) shows that, for $\tau=1$, which corresponds to the situation where there is only global movement, the time-dependent ratio of infected individuals $\rho_I$ has little to do with people's sensitivity to the risk information. Within the range of $0<t<500$, $\rho_I$ decreases quickly from $\rho_I\sim 0.086$ to $\rho_I\sim 0$ for different values of people's sensitivity to the risk information. Within the range of $t>500$, $\rho_I$ keeps the minimum value of $\rho_I\sim 0$. Such results indicate that, in the global person-to-person movement, an individual's self-restricted travel behavior has little effect on the change of the spread of an infectious disease.

Comparing the simulation results in figure 1 (a), (b) and (c), we find that the effectiveness of people's self-restricted travel behaviors on the suppression of an infectious disease is related to the timescale between global movement and local movement. In the local movement, as people can not get exact epidemic information about who has been infected and who has not, people's strict self-restriction is quite important for us to inhibit the spread of an infectious disease. It is valuable that people have a higher level of self-consciousness to the prevention and control of the infectious disease. In the global movement, as each individual has timely and accurate epidemic information about whether his immediate neighbors have been infected or not, he can cut off the connections with the infected individuals in time. Therefore, it is not necessary for him to have a strict self-restricted travel behavior as long as he has already cut off all the connections with the infected individuals. In the coexistence of local movement and global movement, as accurate epidemic information is not always available, people's strict self-restriction is still important for us to inhibit the spread of an infectious disease.

\begin{figure}
\includegraphics[width=6cm]{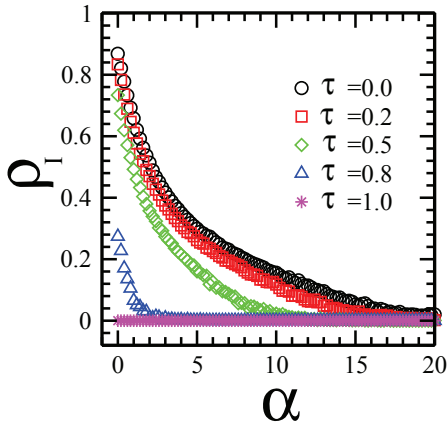}
\caption{\label{fig:epsart}Simulation of averaged ratio of infected individuals $\rho_I$ as a function of people's sensitivity $\alpha$ to the epidemic information with timescale between global and local movement $\tau$=0 (circles), 0.2 (squares), 0.5 (diamonds), 0.8 (triangles), 1 (stars). Other parameters are: $\alpha$=$\beta$, total population $N=1000$,  average degree of each node in the random network $\bar k$=10, initial moving probability $v_0$=1, delayed time $\Delta$=0, infection rate $P_I=1$, recovery rate $P_S=0.01$, global area $L\times L=100\times 100$, local area $l\times l=11\times 11$, overlapping area $l\times\Delta l=11\times1$. Final data are obtained by averaging over 10 runs and $10^3$ time steps after $5\times10^3$ relaxation time in each run.}
\end{figure}

In order to get a clear view of the extinction point of the epidemic, we plot the averaged ratio of infected individuals $\rho_I$ as a function of people's sensitivity $\alpha$ for different values of timescale $\tau$ in Figure 2. As there is only local random movement, $\tau$=0, $\rho_I$ decreases continuously from $\rho_I\sim 0.87$ to $\rho_I\sim 0.02$ within the range of $0<\alpha<20$. The extinction point $\alpha_c\sim 20$ is observed. For $\tau=$0.2, $\rho_I$ decreases continuously from $\rho_I\sim 0.84$ to $\rho_I\sim 0$ within the range of $0<\alpha<20$. The extinction point $\alpha_c\sim 18$ is observed. A further increase in $\tau$ leads to a decrease in the extinction point $\alpha_c$. As there is only global person-to-person movement, $\tau$=1, $\rho_I$ keeps the minimum value of $\rho_I\sim 0$ within the whole range of $0<\alpha<20$.

We can conclude that, if people could get exact epidemic information, just like the situation where there is only global movement, they can effectively refrain from being infected and a strict self-restricted travel behavior has little effect on the change of the spread of the epidemic. If people could only get the ratio of infected individuals in the population, just like the situation where there is only local movement, people's higher sensitivity to the epidemic information would cause them to travel less and the spread of the epidemic can be suppressed effectively.

\begin{figure}
\includegraphics[width=12cm]{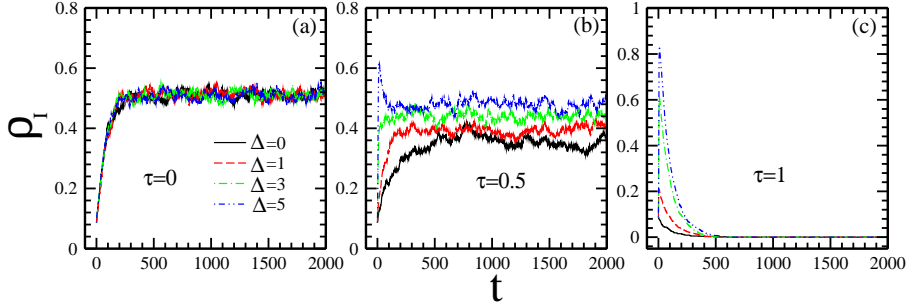}
\caption{\label{fig:epsart}Simulation of dynamic ratio of infected individuals $\rho_I$ for time t with timescale between global and local movement (a) $\tau=0$; (b) $\tau=0.5$; (c) $\tau=1$. The delayed time $\Delta$=0 (lines), 1 (slashes), 3 (slash dotted lines), 5 (slash dotted dotted lines). Other parameters are: total population $N=1000$, average degree of each node in the random network $\bar k$=10, initial moving probability $v_0$=1, people's sensitivity to the epidemic information $\alpha$=$\beta$=2, infection rate $P_I=1$, recovery rate $P_S=0.01$, global area $L\times L=100\times 100$, local area $l\times l=11\times 11$, overlapping area $l\times\Delta l=11\times1$.}
\end{figure}

Secondly, incorporating the delayed epidemic information into the evolutionary process, we examine whether the delayed epidemic information would affect the spread of an infectious disease or not.

Figure 3 (a) shows that, for $\tau=0$, which corresponds to the situation where there is only local movement, the time-dependent ratio of infected individuals $\rho_I$ are nearly the same for different values of the delayed time $\Delta$. Such results indicate that, in the local random movement, whether the epidemic information is delayed or not does not affect the spread of an infectious disease.

Figure 3 (b) shows that, for $\tau=0.5$, which corresponds to the situation where local movement and global movement coexist, the time-dependent ratio of infected individuals $\rho_I$ is related to the delayed time $\Delta$ of the epidemic information. For $\Delta$=0, which corresponds to the situation where people have timely epidemic information, the ratio of infected individuals $\rho_I$ increases slowly from $\rho_I\sim 0.1$ to $\rho_I\sim 0.36$ within the range of $0<t<500$. For $t>500$, $\rho_I$ fluctuates around $\rho_I\sim 0.36$ and the average value of $\rho_I$ changes little with the rise of $t$. For $\Delta$=1, which corresponds to the situation where people have delayed epidemic information, $\rho_I$ increases from $\rho_I\sim 0.1$ to $\rho_I\sim 0.4$ within the range of $0<t<200$. For $t>200$, $\rho_I$ fluctuates around $\rho_I\sim 0.4$ and the average value of $\rho_I$ changes little with the rise of $t$. For $\Delta$=5, $\rho_I$ firstly increases quickly from $\rho_I\sim 0.1$ to $\rho_I\sim 0.62$ and then drops quickly from $\rho_I\sim 0.62$ to $\rho_I\sim 0.48$ within the range of $0<t<150$. For $t>150$, $\rho_I$ fluctuates around $\rho_I\sim 0.48$ and the average value of $\rho_I$ changes little with the rise of $t$.  A further increase in $\Delta$ leads to a higher level of $\rho_I$ in the final steady state and the shortened time to the final steady time. Such results indicate that, in the situation where local movement and global movement coexist, the delayed epidemic information promotes the spread of an infectious disease.

Figure 3 (c) shows that, for $\tau=1$, which corresponds to the situation where there is only global movement, the time-dependent ratio of infected individuals $\rho_I$ is closely related to the delayed time $\Delta$ of the epidemic information. For $\Delta=0$, which corresponds to the situation where people have timely epidemic information, the ratio of infected individuals $\rho_I$ decreases from $\rho_I\sim 0.1$ to $\rho_I\sim 0$ within the range of $0<t<300$. For $t>300$, $\rho_I$ keeps the minimum value of $\rho_I\sim 0$. For $\Delta=1$, which corresponds to the situation where people have delayed epidemic information, the ratio of infected individuals $\rho_I$ firstly increases from $\rho_I\sim 0.1$ to $\rho_I\sim 0.2$ and then decreases from $\rho_I\sim 0.2$ to $\rho_I\sim 0$ within the range of $0<t<500$. For $t>500$, $\rho_I$ keeps the minimum value of $\rho_I\sim 0$. A further increase in $\Delta$ leads to an increase in the maximum value of the time-dependent $\rho_I$. However, the ratio of infected individuals in the final steady does not change with $\Delta$. Such results indicate that, in the situation where there is only global movement, the delayed epidemic information cause a rapid progress and a higher peak of the spread of an infectious disease.

\begin{figure}
\includegraphics[width=8cm]{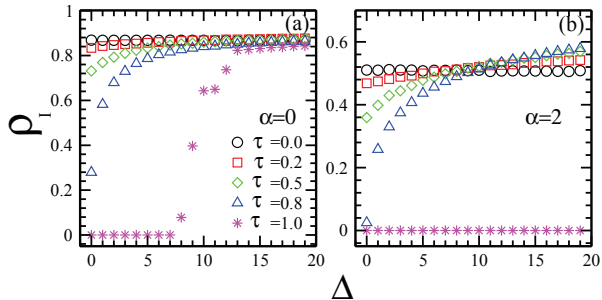}
\caption{\label{fig:epsart}Simulation of averaged ratio of infected individuals $\rho_I$ as a function of delayed time $\Delta$ with (a)people's sensitivity to the epidemic information $\alpha$=0 and timescale between global and local movement $\tau$=0 (circles), 0.2 (squares), 0.5 (diamonds), 0.8 (triangles), 1 (stars); (b)people's sensitivity to the epidemic information $\alpha$=2 and timescale between global and local movement $\tau$=0 (circles), 0.2 (squares), 0.5 (diamonds), 0.8 (triangles), 1 (stars). Other parameters are: $\alpha$=$\beta$, total population $N=1000$, average degree of each node in the random network $\bar k$=10, initial moving probability $v_0$=1, infection rate $P_I=1$, recovery rate $P_S=0.01$, global area $L\times L=100\times 100$, local area $l\times l=11\times 11$, overlapping area $l\times\Delta l=11\times1$. Final data are obtained by averaging over 10 runs and $10^3$ time steps after $5\times10^3$ relaxation time in each run.}
\end{figure}

In order to get a clear view of the coupled effects of the delayed epidemic information and people's self-restricted travel behavior on the spread of an infectious disease, in Figure 4 (a) and (b) we plot the averaged ratio of infected individuals $\rho_I$ as a function of the delayed time $\Delta$ for $\alpha$=0 and $\alpha$=2 respectively.

Figure 4 (a) shows that, for $\alpha$=$\beta$=0, which corresponds to the situation where people are insensitive to the epidemic information, the changing tendency of $\rho_I$ vs $\Delta$ is closely related to the timescale $\tau$. For $\tau=0$, $\rho_I$ keeps its maximum value of $\rho_I\sim 0.88$ within the whole range of $\Delta\ge0$. For $\tau=0.2$, $\rho_I$ increases continuously from $\rho_I\sim 0.83$ to $\rho_I\sim 0.88$ within the range of $0<\Delta<6$. $\rho_I$ keeps its maximum value of $\rho_I\sim 0.88$ within the whole range of $\Delta>6$. The critical value of $\Delta_c\sim6$ is observed. A further increase in $\tau$ leads to an overall decrease in $\rho_I$ within the range of $0<\Delta<\Delta_c$. For $\tau=1$, which corresponds to the situation where there is only global movement, $\rho_I$ keeps its maximum value of $\rho_I\sim 0$ within the range of $0<\Delta<7$. Within the range of $7<\Delta<13$, $\rho_I$ increases quickly from $\rho_I\sim 0$ to $\rho_I\sim 0.84$. Within the range of $\Delta>13$, $\rho_I$ keeps its maximum value of $\rho_I\sim 0.84$.

Figure 4 (b) shows that, for $\alpha$=$\beta$=2, which corresponds to the situation where people are sensitive to the epidemic information, the changing tendency of $\rho_I$ vs $\Delta$ is also closely related to the timescale $\tau$. For $\tau=0$, $\rho_I$ keeps its value of $\rho_I\sim 0.52$ within the whole range of $\Delta>0$. For $\tau=0.2$, $\rho_I$ increases continuously from $\rho_I\sim 0.48$ to $\rho_I\sim 0.54$ within the whole range of $\Delta>0$. A further increase in $\tau$ leads to a decrease in $\rho_I$  within the range of $0<\Delta<\Delta_c$ and an increase in $\rho_I$ within the range of $\Delta>\Delta_c$. For $\tau=1$, which corresponds to the situation where there is only global movement, $\rho_I$ keeps its minimum value of $\rho_I\sim 0$ within the whole range of $\Delta>0$. Comparing the results in Figure 4 (b) with the results in Figure 4 (a), we find that a higher level of people's sensitivity to the epidemic information is helpful for the inhibition of the spread of an infectious disease even if there is a time delay of the epidemic information.

We conclude that, the delayed epidemic information does not always have an impact on the spread of the epidemic. In the situation where there is only local movement, whether the epidemic information is delayed or not does not affect the spread of the epidemic. In the situation where local movement and global movement coexist, the delayed epidemic information promotes the widespread of the epidemic. In the situation where there is only global movement, the delayed epidemic information only causes an increase in the highest level of the spread of the epidemic but does not affect the ratio of the infected individuals in the final steady state.

\begin{figure}
\includegraphics[width=8cm]{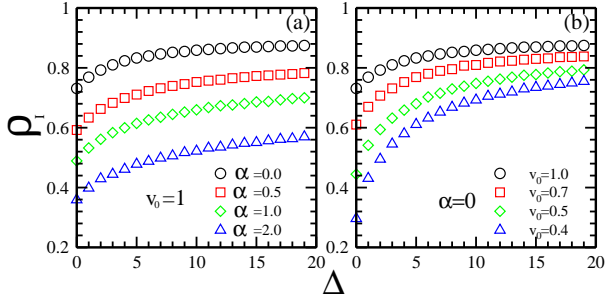}
\caption{\label{fig:epsart}Simulation of averaged ratio of infected individuals $\rho_I$ as a function of delayed time $\Delta$ with (a)initial moving probability $v_0=1$ and people's sensitivity to the epidemic information $\alpha$=0 (circles), 0.5 (squares), 1 (diamonds), 2 (triangles); (b)people's sensitivity to the epidemic information $\alpha$=0 and initial moving probability $v_0$=1 (circles), 0.7 (squares), 0.5 (diamonds), 0.4 (triangles). Other parameters are: $\alpha$=$\beta$, total population $N=1000$, average degree of each node in the random network $\bar k$=10, timescale between global and local movement $\tau=0.5$, infection rate $P_I=1$, recovery rate $P_S=0.01$, global area $L\times L=100\times 100$, local area $l\times l=11\times 11$, overlapping area $l\times\Delta l=11\times1$. Final data are obtained by averaging over 10 runs and $10^3$ time steps after $5\times10^3$ relaxation time in each run.}
\end{figure}

In the present model, people's sensitivity to the epidemic information is reflected in their self-restricted travel behavior. The change in people's travel behavior determines people's encounter probability and accordingly the spread of the epidemic. In real society, the moving probability can be affected by external forces, like government's speed limits, and internal forces, like an individual's risk aversion behavior. In order to get a clear view of different effects of external forces and internal forces on the spread of an infectious disease, in Figure 5 (a) and (b) we plot the averaged ratio of infected individuals as a function of timescale $\tau$ for a fixed value of $v_0=0$ and different values of $\alpha$ and a fixed value of $\alpha=0$ and different values of $v_0$ respectively. The scenario in Figure 5 (a) represents people's risk aversion behaviors, in which people's moving probability changes with time. The scenario in Figure 5 (b) represents government's speed limits, in which people's moving probability keeps low but stable. We are especially concerned about the epidemic inhibition effects between the system with low and stable moving probability and the system with information-dependent moving probability.

Figure 5 (a) shows that, for $v_0=1$, the changing tendencies of $\rho_I$ vs $\Delta$ are nearly the same for different values of people's sensitivity to the epidemic information $\alpha$. An increase in $\alpha$ only leads to an overall decrease in $\rho_I$ but not the changing tendency of $\rho_I$ vs $\Delta$. Figure 5 (b) shows that, for $\alpha=0$, the changing tendencies of $\rho_I$ vs $\Delta$ are not the same for different values of initial moving probability $v_0$. An increase in $v_0$ not only leads to an overall decrease in $\rho_I$ but also the changing tendency of $\rho_I$ vs $\Delta$.

Comparing the results in Figure 5 (a) with the results in Figure 5 (b), we find that the changing tendency of $\rho_I$ vs $\Delta$ in Figure 5 (b)  is much steeper than that in Figure 5 (a). Such results indicate that, compared with the situation where the moving probability keeps low and stable, people's self-restricted travel behavior is more effective in reducing the disadvantage resulting from the delayed epidemic information.

\section{Theoretical analysis}
\label{sec:analysis}

\subsection{\label{subsec:levelB}Relationship between the ratio of infected individuals and people's sensitivity to the epidemic information}

\begin{figure}
\includegraphics[width=5cm]{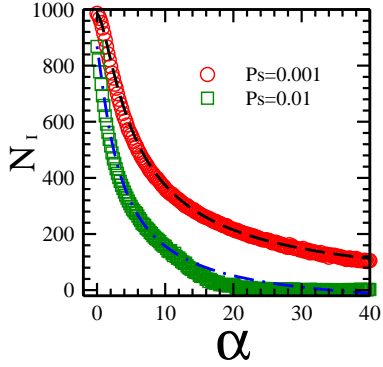}
\caption{\label{fig:epsart}Simulation of the averaged total number of infected individuals $N_I$ as a function of people's sensitivity to the epidemic information $\alpha$ with  recovery rate $P_S$=0.001 (circles), 0.01 (squares). Other parameters are: $\alpha$=$\beta$, total population $N=1000$, average degree of each node in the random network $\bar k$=10, initial moving probability $v_0$=1, timescale between global and local movement $\tau$=0, delayed time $\Delta$=0, infection rate $P_I=1$,  global area $L\times L=100\times 100$, local area $l\times l=11\times 11$, overlapping area $l\times\Delta l=11\times1$. Final data are obtained by averaging over 10 runs and $10^3$ time steps after $5\times10^3$ relaxation time in each run. The slash lines and the slash dotted lines are theoretical predictions for $P_S$=0.001 and 0.01 respectively.}
\end{figure}
Here, we give a theoretical analysis on how people's sensitivity to the epidemic information affects the ratio of infected individuals in the final steady state.

In the present model, an increase or a decrease in the number of infected individuals is determined by the encounter probability between the infected and the suspected individuals. In the following, depending upon the mean field theory, we give an analysis on how an individual's risk aversion behavior affects the encounter probability, which further leads to the change of the ratio of infected individuals in the final steady state.

For $\tau=0$, which corresponds to the situation where there is only local random movement, on condition that people are insensitive to the risk information, $\alpha$=0, each individual's travel probability keeps stable, $v(t)$=$v_0$=1. The encounter probability between an infected individual and a susceptible individual is

\begin{equation}
P_{SI}=\frac{C^1_{N_I}C^1_{N_S}}{C^2_{N}}=\frac{2N_IN_S}{(N-1)(N-2)},
\end{equation}
in which $N_I$ is the number of infected individuals and $N_S$ is the number of susceptible individuals in the population and $N_I$+$N_S$=N. Suppose the relationship between $P_{SI}$ and $v_0$ is satisfied with a linear function,

\begin{equation}
P_{SI}=\frac{2N_I(N-N_I)v_0}{(N-1)(N-2)},
\end{equation}
an increase in the ratio of infected individuals in a time step is

\begin{equation}
\Delta \rho_I=P_{SI}=\frac{2N_I(N-N_I)v_0}{(N-1)(N-2)}.
\end{equation}

As people's sensitivity to the risk information is incorporated into the travel probability, $v=v_0(1-\frac{N_I}{N})^\alpha$, $\Delta \rho_I$ becomes

\begin{equation}
\Delta \rho_I=\frac{2N_I(N-N_I)v_0(1-\frac{N_I}{N})^\alpha}{(N-1)(N-2)}.
\end{equation}

With the recovery probability $P_S$, a decrease in the ratio of infected individuals in a time step is

\begin{equation}
\Delta \rho'_I=\frac{P_SN_I}{N}.
\end{equation}

In the final stable state, an increase in the ratio of infected individuals should be equal to a decrease in the ratio of infected individuals, $\Delta \rho_I=\Delta \rho'_I$, we get

\begin{equation}
\frac{2N_I(N-N_I)^{1+\alpha}v_0}{(N-1)(N-2)N^\alpha}=\frac{P_SN_I}{N}.
\end{equation}
The relationship between the number of infected individuals in the population and people's sensitivity to the risk information becomes

\begin{equation}
N_I=N-[\frac{(N-1)(N-2)N^{\alpha-1} P_S}{2v_0}]^{\frac{1}{1+\alpha}}.
\end{equation}

In the above equation, there is a pre-defined condition that the population density is $\rho=1$. Consider different population density $\rho$, we get the functional relation

\begin{equation}
N_I=N-[\frac{(N-1)(N-2)N^{\alpha-1} P_S}{2\rho v_0}]^{\frac{1}{1+\alpha}}.
\end{equation}

From the above equation we find that the amount of infected individuals in the final steady state decreases with the rise of the recovery probability and the rise of people's sensitivity to the risk information. In Figure 6 we plot $N_I$ vs $\alpha$ for different $P_S$. Given the condition $N^{theo}_I(\alpha=0)=N^{simul}_I(\alpha=0)$, the theoretical analysis is in accordance with the simulation data.

\subsection{\label{subsec:levelB}relationship between the progress of the epidemic and the delayed epidemic information}

\begin{figure}
\includegraphics[width=5cm]{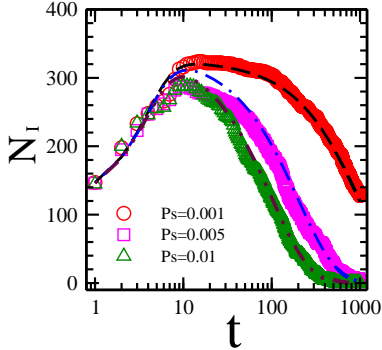}
\caption{\label{fig:epsart}Simulation of the dynamic total number of infected individuals $N_I$ for time t with recovery rate $P_S$=0.001 (circles), 0.005 (squares), 0.01 (triangles). Other parameters are: total population $N=1000$, average degree of each node in the random network $\bar k$=10, initial moving velocity $v_0$=1, timescale between global and local movement $\tau=1$, delayed time $\Delta$=1, people's sensitivity to the epidemic information $\alpha$=$\beta$=0, infection rate $P_I=1$, global area $L\times L=100\times 100$, local area $l\times l=11\times 11$, overlapping area $l\times\Delta l=11\times1$. The slash lines, the slash dotted lines and the slash dotted dotted lines are theoretical predictions for $P_S$=0.001, 0.005 and 0.01 respectively.}
\end{figure}

In the following, we give a theoretical analysis on how the delayed epidemic information affects the progress and the peak of an epidemic.

In the local random movement, an individual's travel probability is determined by the ratio of infected individuals in the local area. According to the mean field theory, the ratio of infected individuals in the local area is equal to the ratio of infected individuals in the whole area, which is somewhat stable in the final steady state. Therefore, although the state of each individual changes with time, the delayed epidemic information also provides us relatively accurate information of $\rho_I$ and does not affect the progress of an epidemic.

In the global person-to-person movement, an individual's travel probability is determined by accurate epidemic information about who has been infected and who has not. If an individual has accurate epidemic information, his travel probability is

\begin{equation}
v_{global}=\frac{n-n_I}{n},
\end{equation}
in which $n_I$ is the number of infected individuals and $n$ is the total number of individuals connected to him. In such cases, a susceptible individual's connections with infected individuals have all been cut off and he will not get infected.

If the epidemic information is delayed one time step, an individual's travel probability becomes

\begin{equation}
v_{global}=\frac{n-(n_I-\Delta n_I)}{n},
\end{equation}
in which $\Delta n_I$ is the number of infected individuals increased in the latest step. Depending upon the mean field theory, we get the possibility that an individual will get infected in the next time step

\begin{equation}
\Delta\rho_I=\frac{C^1_{N_S}C^1_{\Delta N_I}}{C^2_N},
\end{equation}
in which $N_S$ is the number of susceptible individuals and $\Delta N_I$ is the number of infected individuals increased in the latest step in the whole population. We obtain the number of infected individuals increased in a time step

\begin{equation}
\Delta N_I(t)=\frac{2N^2_S(t-1)\Delta N_I(t-1)}{N(N-1)}.
\end{equation}
Therefore, the total number of infected individuals is

\begin{equation}
N_I(t)=N_I(t-1)+\frac{2N^2_S(t-1)\Delta N_I(t-1)}{N(N-1)}-N_I(t-1)P_S,
\end{equation}
and the total number of susceptible individuals is

\begin{equation}
N_S(t)=N_S(t-1)-\frac{2N^2_S(t-1)\Delta N_I(t-1)}{N(N-1)}+N_I(t-1)P_S.
\end{equation}

From the above equations we find that the number of infected individuals in the population firstly increases and then decreases with time, the rise or the decrease slope of which is closely related to the refractory probability $P_S$. In figure 7 we plot $N_I$ vs $t$ for different $P_S$. Given the initial conditions $N_I=112$, $N_S=795$, $\Delta N_I=30$, $\Delta N_S=0$, the theoretical analysis is in accordance with the simulation data.

\section{Summary}

In this paper, we have investigated the coupled effects of delayed epidemic information and self-restricted travel behavior on contagion. In the local random movement, people's self-restricted travel behavior can effectively reduce the spread of an infectious disease. Compared with timely epidemic information, delayed epidemic information does not reduce the effectiveness of such an advantage. In the global person-to-person movement,  people's self-restricted travel behavior is also helpful for us to prevent and control the epidemic. The effectiveness of such an advantage depends on timely and accurate epidemic information. The delayed epidemic information facilitates the epidemic reaching a higher peak. In the coexistence of local movement and global movement, both timely epidemic information and people's high sensitivity to the risk information are helpful for us to suppress the epidemic. A mean field analysis indicates that, in the local random movement, the suppression of the epidemic results from a decrease in the encounter probability between the susceptible and the infected. People's self-restricted travel behavior leads to a lower Susceptible-Infected encounter probability and the ratio of infected individuals decreases. In the global person-to-person movement, the suppression of the epidemic results from timely and accurate disconnection of the link between the susceptible and the infected. The timely and accurate epidemic information helps an individual know of who has been infected and who has not, depending upon which he can cut off the Susceptible-Infected linkage in time and the epidemic is suppressed.

In real society, there are a variety of moving patterns. From the perspective of effectively curbing an epidemic, an infectious disease's transmission characteristics and people's moving patterns are two key factors, depending upon which we can map out more effective prevention and control measures. Compared with external enforcement measures, people's self-restricted travel behavior are more flexible and helpful, which help us achieve a balance between the prevention and control of the epidemic and the needs of outdoor activities in our daily life. In the future, a challenging problem is how to give a detailed theoretical derivation of a typical kind of coupled reaction-diffusion processes, including a comprehensive consideration of information transmission factors and cognitive factors. A kind of message-passing method may be a promising protocol for such problems.

\section*{Acknowledgments}

This work is the research fruits of Social Science Foundation of Zhejiang Province, National Social Science Foundation of China, Humanities and Social Sciences Fund sponsored by Ministry of Education of China (Grant Nos. 19YJAZH120, 17YJAZH067), National Natural Science Foundation of China (Grant Nos. 71371165, 11865009, 71871094, 71631005).





\bibliographystyle{model1-num-names}



\end{document}